\documentclass[12pt]{article}
\usepackage{amsmath}
\usepackage{amssymb}
\usepackage{graphicx}
\usepackage{float}
\usepackage{indentfirst}
\usepackage{mathrsfs}

\usepackage{setspace}

\usepackage[labelfont=bf,labelsep=period]{caption}

\usepackage[numbers,sort&compress]{natbib}

\newtheorem{theorem}{Theorem}
\newtheorem{proposition}{Proposition}
\newtheorem{corollary}{Corollary}

\title{Equivalence Theorem of Uncertainty Relations}

\author
{Jun-Li Li$^{1,2}$ and Cong-Feng Qiao$^{1,2,3\ast}$\\ [0.2cm]
\normalsize{$^1$Department of Physics, University of the Chinese Academy of Sciences,}\\
\normalsize{YuQuan Road 19A, Beijing 100049, China}\\[2pt]
\normalsize{$^2$Key Laboratory of Vacuum Physics, University of Chinese Academy of Sciences}\\[2pt]
\normalsize{$^3$CAS Center for Excellence in Particle Physics, Beijing 100049, China}\\[3mm]
\normalsize{$^\ast$To whom correspondence should be addressed;
E-mail:  qiaocf@ucas.ac.cn.}
}

\date{}

\begin{document}
\baselineskip24pt \maketitle

\begin{abstract}\doublespacing
We present an equivalence theorem to unify the two classes of uncertainty relations, i.e., the variance-based ones and the entropic forms, which shows that the entropy of an operator in a quantum system can be built from the variances of a set of commutative operators. That means an uncertainty relation in the language of entropy may be mapped onto a variance-based one, and vice versa. Employing the equivalence theorem, alternative formulations of entropic uncertainty relations stronger than existing ones in the literature are obtained for qubit system, and variance based uncertainty relations for spin systems are reached from the corresponding entropic uncertainty relations.
\end{abstract}

\section{Introduction}

The renowned uncertainty principle is one of the distinctive features of quantum mechanics, which was introduced by Heisenberg in the description of microscopic quantum behavior \cite{Heisenberg-o}, somewhat similar to the concept of complementary raised by Bohr \cite{Complementary-o}. The uncertainty relation(UR) is a mathematical expression for uncertainty principle, referring to the repulsive nature of incompatible operators and hence imposing a strong restriction on the outcomes of any joint measurement on those operators. Since the UR has profound influence on various aspects of quantum information, e.g. quantum nonlocality \cite{correlation-1,correlation-2,correlation-3}, entanglement \cite{Ent-det}, and quantum cryptography \cite{crypto}, the study on it has never stopped.

It is well-known that the most famous UR, the Heisenberg-Robertson UR \cite{Robertson}, bounds the product of the variances of observables $A$ and $B$ through
the expectation value of their commutator, i.e.,
\begin{eqnarray}
\Delta A\Delta B\geq |\langle C\rangle|\;  \label{Robertson-U}\ .
\end{eqnarray}
Here, the state $|\psi\rangle$ is arbitrary, $\Delta
X = \sqrt{\langle\psi| X^2|\psi\rangle - \langle\psi| X|\psi\rangle^2}$ is
the square root of the variance of a given operator $X$, and $[A,B] = 2iC$
is the commutator of operators $A$ and $B$. Note, the relation (\ref{Robertson-U}) is applicable to any pairs of operators, rather than only to conjugate observables. Though later improvement \cite{Schrodinger-uncertainty,psi-perp} strengthened the UR, the state-dependent feature of lower bound remains, which implies the null lower bound  triviality
\cite{Deutsch-uncertainty}. Early attempts in searching the
state-independent lower bounds led only to near-optimal results
\cite{V-E-Huang}. In a recent work \cite{Reformulating}, this triviality problem was solved at length, by which the state-independent optimal trade-off relations for the variances of multiple observables are found to be obtainable, at least in principle is, see also \cite{tight-variance} for further development of URs involving variances of multiple observables.

It was noticed that the variance is inadequate in quantifying uncertainty,
i.e. relabeling of the non-degenerate eigenvalues of an operator may alter the value of its variance \cite{Deutsch-uncertainty}. To overcome this problem,
the concept of entropy was employed, and a typical
UR in entropy tells \cite{entropy-MU}
\begin{eqnarray}
H(A)+H(B)\geq -2\ln c_{ab} \; . \label{entropy-mu}
\end{eqnarray}
Here, $H(A) = -\sum_j p_j\ln p_j$ is the Shannon entroy, with $p_j$ the
probability distribution of the eigenbasis $\{|a_j\rangle\}$ of operator $A$
in measuring system, and similarly for $H(B)$. The bound $c_{ab} =
\mathrm{Max}_{j,k}|\langle a_j|b_k\rangle|$ is the maximum overlap of
eigenbases of operators $A$ and $B$ and independent of the quantum state. To
construct an optimal entropic UR, the key point is to find
the best lower bound, which usually is a tough question for general
observables \cite{entropy-VS,entropy-Comment} and the optimal bounds for
qubit system with Shannon \cite{optimal-1,optimal-2,optimal-3} and collision
entropies \cite{optimal-4} were obtained. We refer to \cite{Zozor-qubit,Coles-Piani,Strong-major,Entropy-fun, entropy-anti} for the more involved situations of this problem. The entropic uncertainty relation may in principle
apply to multiple observables, of which in fact the state-independent lower
bounds have been investigated for mutually unbiased bases
\cite{MUB-1,MUB-2}, see Refs. \cite{MUB-rev,Entropy-rev} for recent reviews.

There are actually many types of entropies capable of characterizing the
quantum uncertainty \cite{Zozor-qubit}, e.g., R\'enyi entropies
$H_{\alpha} (A) =\ln(\sum_j p_j^{\alpha})/(1-\alpha)$ with different
indices of positive real numbers $\alpha \in \mathbb{R}^+$ (when $\alpha\to 1$ it is Shannon entropy). As there is no obvious reason why one type of entropy is superior to others in the context of UR, a new characterization of uncertainty was introduced: the majorization of the probability distribution \cite{universal-probabilities, Majorization-1, Strong-major}, which is closely related to the entropy.
Since both variance and entropy originate from the probability distribution
of measurement, one may naturally ask: are these two types of uncertainty
relations relevant or equivalent? We find in this work that the answer is definite.

In the following we will present a general scheme on how to build
quantitative relations between two prominent uncertainty measures, the
variance and the entropy, which indicates that those two classes of
URs may actually be unified. The scheme can be sketched as
follows: first construct a set of commutative operators for a given physical
observable, then reconstruct the probability distribution of the measurement
outcomes of the physical observable from the variances of operators in the
set. Based on the quantitative relation, we get various entropic URs for qubit system from variance-based URs, which give optimal lower and upper bounds for arbitrary entropic measures and for multiple observables beyond mutually unbiased bases. Moreover, new variance-based URs for high spin systems are also obtained from the entropic URs.

\section{The equivalence theorem and its applications}

In quantum mechanics, a physical system can be generally described by
density operator $\rho$, which is a positive definite Hermitian matrix; and
a physical observable is represented by a Hermitian operator and may be
expressed through the spectral decomposition $A = \sum_{j=1}^N \lambda_j
|j\rangle \langle j|$, where $|j\rangle$ are the eigenvectors of $A$ with
the corresponding eigenvalues $\lambda_j$. When measuring observable $A$ in
a quantum system $\rho$, only its eigenvalues $\lambda_j$ are attainable
with certain probability in every individual measurement. In
$\rho$ ensemble, the probability of measuring $\lambda_j$ reads $p_j=\langle
j|\rho |j\rangle$. This statistical interpretation leads to two uncertainty
measures, the variance and entropy, which are mathematically expressed as
\begin{eqnarray}
& V(A) & = \ \frac{1}{2}\sum_{j,k=1}^N p_jp_k(\lambda_j-\lambda_k)^2 \; , \label{V-g} \\
& H_{\alpha}(A) &\, = \ \frac{1}{1-\alpha}\ln(\sum_{j=1}^N p_j^{\alpha}) \; . \label{H-a}
\end{eqnarray}
Here, $V(A)$ signifies variance defined as $V(A) \equiv \Delta A^2 =
\mathrm{Tr}[\rho A^2]-\mathrm{Tr}[\rho A]^2$, and $H_{\alpha}(A)$ represents the R\'enyi entropy. We deal with R\'enyi entropy through this article, and for general forms of entropic functions we refer to \cite{Entropy-fun}. Notice that subtracting a constant from the operator does not change its variance and entropy, we are hence legitimate to treat the operators in following discussion to be traceless.

\subsection{The equivalence theorem}

Following we exhibit an equivalence theorem, the main result of this work, which may quantitatively relate different uncertainty measures of discrete systems.
\begin{theorem}
For a given physical observable $A$ in N-dimensional representation with
eigenbases $|j\rangle$, there exists a set of commutative operators,
$\mathcal{A} = \{A_i| A_i = \sum_{j=1}^N \lambda^{(i)}_j|j\rangle\langle
j|\; , A_1 = A \}$, whose variances in quantum state $\rho$ are
\begin{equation}
\Delta A_i^2 = \sum_{k>j=1}^{N}p_jp_kg_{jk}^{(i)} \; , \;
\text{with}\ g_{jk}^{(i)} = (\lambda_j^{(i)}-\lambda_k^{(i)})^2 \; , \label{theorem-eq}
\end{equation}
from which the probability distribution $p_j = \langle j|\rho|j\rangle$
could be uniquely determined. Here $A = A_1\in \mathcal{A}$, and the infimum of the cardinality of the set
$\mathcal{A}$ lies in $[N-1, N(N-1)/2]$.
\end{theorem}

\noindent {\bf Proof:}  Let $l=(j-1)N+k-(j+1)j/2$, then there is an one-to-one correspondence between integer $l$ and the integer array $(j,k)$ and Eq.
(\ref{theorem-eq}) may be rewritten as
\begin{equation}
\Delta A_i^2\ = \sum_{l=1}^{N(N-1)/2} G_{il} \, x_l \; , \label{new-variable}
\end{equation}
where $G_{il} =g_{jk}^{(i)}$ and $x_l = p_jp_k$ with $k>j$. The number of
linear equations (\ref{new-variable}) equals to the cardinality of the set
$\mathcal{A}$ which we denote as $|\mathcal{A}|$. When $|\mathcal{A}| =
N(N-1)/2$, the coefficient matrix $G_{il}$ can be constructed to be invertible by assigning specific values to $\lambda_{j}^{(i)}$ for $i=1, 2, \cdots, N(N-1)/2$. The solutions of $x_l$ are linear functions of $\Delta A_i^2$, which in
turn yields $N(N-1)/2$ equations for $p_j$
\begin{eqnarray}
 p_jp_k =x_l(\Delta A_1^2,\cdots,\Delta A^2_{N(N-1)/2})\; . \label{over-determined}
\end{eqnarray}
Here $\Delta A_i^2$ are function arguments of $x_l(\cdot)$, from which $p_j$ can also be uniquely determined as functions of $\Delta A_i^2$.

As Eq. (\ref{over-determined}) is an over determined equation system, we need not to know all the $N(N-1)/2$ variables of $x_l$ to
uniquely determine the $N$ variables $p_j$ . That means the set $\mathcal{A}$ may be even constructed with $|\mathcal{A}| \leq N(N-1)/2$. On the other hand, the number of equations constraining $p_j$ cannot be less than $N$, otherwise the solution of $p_j$ will not be unique. Considering the additional constraint $\sum_{j=1}^Np_j=1$, $|\mathcal{A}|$ must be greater than or equal to $N-1$, the dimension of Cartan subalgebra of SU($N$) group. In all, the cardinality of the set $|\mathcal{A}|$ lies in $[N-1, N(N-1)/2]$. Q.E.D.

Theorem 1 applies for arbitrary physical observables. When the observable is non-degenerate, i.e., $\forall i\neq j$, $\lambda_i\neq \lambda_j$, the commutative set $\mathcal{A}$ could be constructed explicitly and the following proposition holds.
\begin{proposition}
For non-degenerated observable $A$ in $N$-dimensional representation with eigenbases $|i\rangle$, the probability distribution $p_i = \langle i|\rho|i\rangle$ in a quantum state $\rho$ may be expressed in terms of covariance functions
\begin{eqnarray}
p_i^2 = \frac{\Omega_{ij}\Omega_{ik}}{\Omega_{jk}} \; , \label{p-poly-result}
\end{eqnarray}
where $\Omega_{ij} \equiv -\mathrm{cov}(\ell_i,\ell_j), 1\leq i<j\leq N$, with the covariance function $\mathrm{cov}(\ell_i,\ell_j) = \langle \ell_i(A) \ell_j(A)\rangle - \langle \ell_i(A)\rangle \langle \ell_j(A)\rangle$, and the Lagrange basis polynomials $\ell_j(x)=\displaystyle \prod_{\substack{m=1 \\m\neq j}}^N \frac{x-\lambda_m}{\lambda_j-\lambda_m}$. \label{Proposition-1}
\end{proposition}

\noindent {\bf Proof:} The least degree polynomial function, that assumed to be valued as $f(\lambda_i)$ for $N$ distinct $\lambda_i$, is a linear combination of Lagrange basis polynomials $f(x) = \sum_{j=1}^N f(\lambda_j) \ell_{j}(x)$, where $\ell_j(x)=\displaystyle \prod_{\substack{m=1 \\m\neq j}}^N \frac{x-\lambda_m}{\lambda_j-\lambda_m}$. The variance of the operator function $f(A)$ may be expressed as
\begin{eqnarray}
\Delta f(A)^2 = \sum_{k>j=1}^N p_jp_k[f(\lambda_j) - f(\lambda_k)]^2 \; ,
\end{eqnarray}
according to Eq.(\ref{V-g}). By setting $f(\lambda_1)-f(\lambda_k) = \alpha_{k-1}$, we have
\begin{equation}
\Delta f(A)^2 = \sum_{j=2}^{N}p_1p_j\alpha_{j-1}^2 + \sum_{k>j=2}^N(\alpha_{j-1}-\alpha_{k-1})^2p_{j}p_{k} \; . \label{p-cov-alpha}
\end{equation}
On the other hand, the function $\widetilde{f}(x) \equiv f(x) - f(\lambda_1)$ has the values $f(\lambda_i)-f(\lambda_1)=-\alpha_{i-1}$ for $N$ distinct $\lambda_i$, therefore $\widetilde{f}(x) = -\sum_{i=2}^N\alpha_{i-1}\ell_i(x)$.  Because $\Delta f(A)^2 = \Delta \widetilde{f}(A)^2$, we have
\begin{eqnarray}
\Delta f(A)^2 & = & \sum_{i,j=2}^N \alpha_{i-1} \alpha_{j-1} ( \langle \ell_i(A)\ell_j(A) \rangle - \langle \ell_i(A) \rangle \langle \ell_j(A) \rangle ) \nonumber \\
& = & \sum_{i=2}^N \alpha_{i-1}^2\mathrm{cov}(\ell_i,\ell_i) \nonumber \\
& & + \sum_{n>m=2}^N [(\alpha_{m-1}^2+\alpha_{n-1}^2)-(\alpha_{m-1}-\alpha_{n-1})^2] \mathrm{cov}(\ell_m,\ell_n) \nonumber \\
& = & \sum_{i=2}^{N} \alpha_{i-1}^2 \sum_{j=2}^{N} \mathrm{cov}(\ell_{i}, \ell_j) - \sum_{n>m=2}^{N} (\alpha_{m-1}-\alpha_{n-1})^2 \mathrm{cov}(\ell_{m},\ell_{n}) \; . \label{v-cov-alpha}
\end{eqnarray}
Here $\mathrm{cov}(\ell_i,\ell_j) = \langle \ell_i(A) \ell_j(A)\rangle - \langle \ell_i(A) \rangle \langle \ell_j(A)\rangle$. The expectation value $\langle \ell_i(A) \rangle = \mathrm{Tr}[\rho \ell_i(A)]$ when mixed states are involved.
The equivalence of Eqs. (\ref{p-cov-alpha}) and (\ref{v-cov-alpha}) does not depend on the values of $\alpha_i$, hence
\begin{equation}
p_1p_j = \sum_{k=2}^N \mathrm{cov}(\ell_j,\ell_k) \, ; \,
p_{j}p_k = -\mathrm{cov}(\ell_j,\ell_k) \; , \; k>j\geq 2\; .
\end{equation}
Using that $\sum_{i}\ell_i(x) = 1$ we have $p_ip_j = -\mathrm{cov}(\ell_i,\ell_j)$, for $i<j$, and Eq. (\ref{p-poly-result}) is obtained. Q.E.D.

Proposition 1 gives the most direct relations between the probabilities and covariances which inherit from the characteristic functions that relates probability distribution and high-order momentums in probability theory. Next, we shall illustrate the extraordinary function of the equivalence theorem in bridging the prevailing variance-based and entropic URs through concrete examples of spin systems.

\subsection{Uncertainty relations for qubits}

Qubit system might be the mostly investigated system in quantum information,
which possesses enormous potential in application. In such systems, any
physical observable may be represented by a $2\times 2$ traceless Hermitian
matrix, and therefore the eigenvalues of an operator may be assigned as
$\lambda_2=-\lambda_1=\lambda$. According to Proposition 1 the following
corollary holds.
\begin{corollary}
In a qubit system, there exists the following monotonic functional relations
between the entropy and the variance
\begin{eqnarray}
H_{\alpha}(A) & = & f_{\alpha}(\Delta A^2) =
\frac{1}{1-\alpha}\ln(a_+^{\alpha}
+a_-^{\alpha}) \; , \label{VtoH} \\
\Delta A^2 \hspace{0.2cm} & = & f_{\alpha}^{-1}[H_{\alpha}(A)]
\equiv g_{\alpha}(A) \; , \label{HtoV-qubit}
\end{eqnarray}
where $a_{\pm}\equiv(1\pm\sqrt{1-\Delta A^2})/2$ with the eigenvalues of A being absorbed into its variance $\Delta A^2/\lambda^2 \to \Delta A^2$, and $f^{-1}_{\alpha}$ is the inverse function
of $f_{\alpha}$. \label{corollary-qubit}
\end{corollary}

\noindent{\bf Proof:} For qubit system where $N=2$, we have $\ell_1(A) =(A-\lambda_2)/(\lambda_1-\lambda_2)$, $\ell_2(A) =(A-\lambda_1)/(\lambda_2-\lambda_1)$, and
\begin{eqnarray}
p_1p_2  = -\mathrm{cov}(\ell_1,\ell_2) = \frac{\Delta A^2}{4\lambda^2} \; .
\end{eqnarray}
Absorbing $\lambda^2$ into $\Delta A^2$, and considering of $p_1+p_2=1$, then
\begin{eqnarray}
p_1 = \frac{1+\sqrt{1-\Delta A^2}}{2} \; ,\;
p_2 = \frac{1-\sqrt{1-\Delta A^2}}{2} \; . \label{probability-2}
\end{eqnarray}
Substituting Eq. (\ref{probability-2}) into the definition of R\'enyi entropy
Eq. (\ref{H-a}), we have
\begin{eqnarray}
H_{\alpha}(A) = f_{\alpha}(\Delta A^2) =
\frac{1}{1-\alpha}\ln(a_+^{\alpha}+a_-^{\alpha}) \; , \label{cor-1eq}
\end{eqnarray}
where $a_{\pm}\equiv(1\pm\sqrt{1-\Delta A^2})/2$. Eq. (\ref{cor-1eq}) is a monotonic function for $\Delta A^2 \in [0,1]$, and therefore
\begin{eqnarray}
\Delta A^2 = f_{\alpha}^{-1}[H_{\alpha}(A)]
\equiv g_{\alpha}(A) \; .
\end{eqnarray}
Here $f^{-1}_{\alpha}$ is the inverse function of the R\'enyi entropy with
index $\alpha$. Q.E.D.

Corollary \ref{corollary-qubit} predicts that, an entropic UR may be converted into a variance-based UR straightforwardly. For example, putting Eq. (\ref{probability-2}) into the entropic UR Eq. (\ref{entropy-mu}) we get
\begin{eqnarray}
a_+^{a_+} \cdot a_-^{a_-} \cdot b_+^{b_+} \cdot b_-^{b_-}\leq c_{ab}^2 \; ,
\end{eqnarray}
where the quantities $a_{\pm} = (1\pm\sqrt{1-\Delta A^2})/2$, $b_{\pm} = (1\pm\sqrt{1-\Delta B^2})/2$.
There is also the majorized UR \cite{universal-probabilities}, $\vec{p}(\rho) \otimes \vec{q}(\rho) \prec \vec{\omega}$, where $\vec{p}(\rho)$ and $\vec{q}(\rho)$ are probability distributions for two observables in quantum state $\rho$, and $\vec{\omega}$ is a state independent vector. Taking Eq. (\ref{probability-2}) into this majorized UR we have
\begin{equation}
(1+\sqrt{1-\Delta A^2})(1+\sqrt{1-\Delta B^2}) \leq (1+c_{ab})^2\; . \label{maj}
\end{equation}
Here $c_{ab}$ is defined in Eq. (\ref{entropy-mu}).

On the other hand, the variance-based UR may also be transformed into entropic UR. However, the state-dependence of the lower bounds of the variance-based URs leads to trivial entropy relations, and the non-trivial results only exist for state-independent ones. For example, we have the variance-based UR from the Theorem 1 of Ref. \cite{Reformulating}, $[a^2(p^2-1)+\Delta A^2] [b^2(p^2-1)+\Delta B^2] \geq (\sqrt{a^2-\Delta A^2}\sqrt{b^2-\Delta B^2} - \kappa p^2 )^2$. Taking Eq. (\ref{HtoV-qubit}) into this variance-based UR we have
\begin{eqnarray}
& & [a^2(p^2-1)+g_{\alpha}(A)] [b^2(p^2-1)+g_{\beta}(B)]
\geq \nonumber \\
& & [ \sqrt{a^2-g_{\alpha}(A)} \sqrt{b^2-g_{\beta}(B)}
- \kappa p^2 ]^2 \; . \label{HtoV-full}
\end{eqnarray}
Here, $a^2 =\mathrm{Tr}[A^2]/2$, $b^2 = \mathrm{Tr}[B^2]/2$, $p^2 =
2\mathrm{Tr}[\rho^2]-1$, and $\kappa = \mathrm{Tr}[AB]/2$; $\alpha$ and
$\beta$ are independent R\'enyi indices. Eq. (\ref{HtoV-full}) gives
both the optimal lower and upper bounds for arbitrary entropic measures, and is tight: Eq. (\ref{HtoV-full}) is satisfied for all the quantum states; for all the values of entropies of operators $A$, $B$ satisfying Eq. (\ref{HtoV-full}), there is the quantum state corresponding to them. This provides the better analytic result comparing to existing ones \cite{optimal-1,optimal-2,optimal-3,optimal-4,Zozor-qubit}. To show this more explicitly, we take pure quantum system with operators $A =\vec{\sigma}\cdot \vec{n}_a$, $B = \vec{\sigma}\cdot \vec{n}_b$ and Shannon
entropies of $\alpha=\beta=1$ as an example. In this case, Eq. (\ref{HtoV-full}) becomes
\begin{equation} g_{1}(A)g_{1}(B) \geq
[\sqrt{1-g_{1}(A)}\sqrt{1-g_{1}(B)} - \cos\theta_{ab}]^2 \;
,\label{HtoV-simple}
\end{equation}
where $\theta_{ab}$ is the angle between unit vectors $\vec{n}_a$,
$\vec{n}_b$. Fig. 1 illustrates the allowed regions for the Shannon
entropies of operators $A$ and $B$ predicted by Eq. (\ref{HtoV-simple}). These figures are consistent with the recent results obtained by analyzing the parameters of state space of qubit \cite{numerical-entropy}.

\begin{figure}[t]\centering
\scalebox{0.25}{\includegraphics{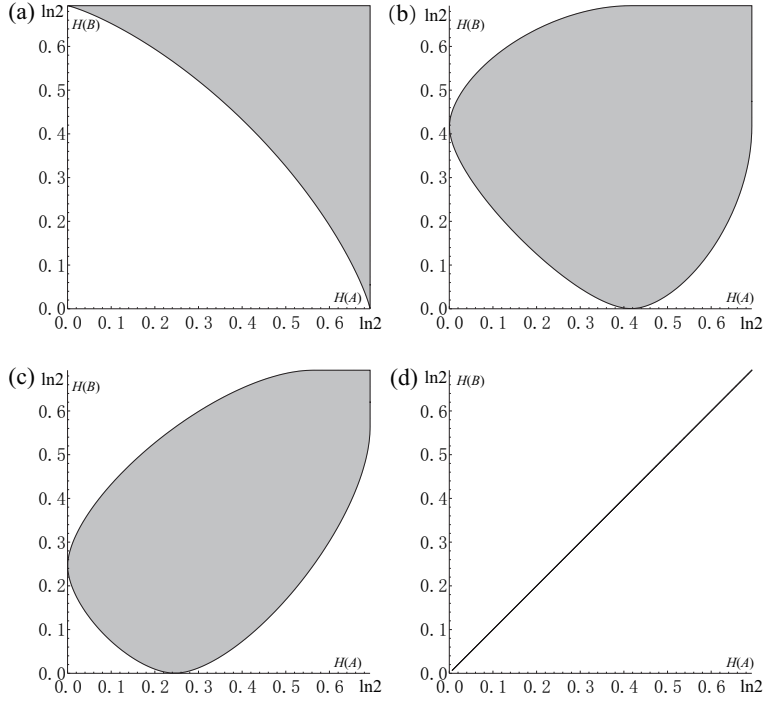}}
\caption{\small The allowed regions for the Shannon entropies of
operators $A=\vec{\sigma}\cdot \vec{n}_a$ and $B=\vec{\sigma}\cdot
\vec{n}_b$ in pure states with (a) $\theta_{ab} = 90^{\circ}$, (b)
$\theta_{ab} = 45^{\circ}$, (c) $\theta_{ab} = 30^{\circ}$, (d)
$\theta_{ab}=0^{\circ}$ respectively.  Here, $\theta_{ab}$ is the angle
between unit vectors $\vec{n}_a$ and $\vec{n}_b$. The obtained entropic
uncertainty relation is optimal: 1. for every point in the shaded area there
is a quantum state that gives the corresponding values of $H(A)$ and $H(B)$;
2. for every quantum state, the values of $H(A)$ and $H(B)$ lie in the
shadow region.}
\end{figure}

For observables more than two, the following corollary exists:
\begin{corollary}
In a qubit system, for three independent observables $A=\vec{\sigma}\cdot
\vec{n}_a$, $B=\vec{\sigma}\cdot \vec{n}_b$, and $C=\vec{\sigma}\cdot
\vec{n}_c$, where $\vec{n}_a$, $\vec{n}_b$, and $\vec{n}_c$ are not coplane,
the entropic UR involving $H_{\alpha}(A)$, $H_{\beta}(B)$,
and $H_{\gamma}(C)$ where $\alpha,\beta,\gamma \in \mathbb{R}^+$, takes the
form of equality.
\end{corollary}
Taking Eq. (\ref{HtoV-qubit}) into Proposition 1 of Ref.
\cite{Reformulating}, one may easily notice that the Corollary 2 holds, and
the equality form of entropic URs could also be obtained explicitly from the variance-based URs for multiple observables \cite{tight-variance}. As an illustration, we take Pauli operators of qubit
system as an example. The variance based uncertainty equality $\Delta
\sigma_x^2 + \Delta \sigma_y^2 + \Delta \sigma_z^2 = 4-2\mathrm{Tr}[\rho^2]$ ( see Refs. \cite{Reformulating,tight-variance}) leads to the following entropic uncertainty
equality
\begin{eqnarray}
g_{\alpha}(\sigma_x) + g_{\beta}(\sigma_y) + g_{\gamma}(\sigma_z)
= 4-2\mathrm{Tr}[\rho^2] \; . \label{3-sigma}
\end{eqnarray}
Here, the function of entropy $g_{\alpha}$ is defined in Eq.
(\ref{HtoV-qubit}). This gives out an optimal equality form of trade-off relations for $H_{\alpha}(\sigma_x)$, $H_{\beta}(\sigma_y)$, $H_{\gamma}(\sigma_z)$ in
arbitrary qubit state, while results given in Refs. \cite{MUB-2, 3sigma-1,
3sigma-existing} provide upper and/or lower bounds in the special
case of $\alpha=\beta=\gamma=1$. For the collision entropy with $\alpha=\beta=\gamma=2$, Eq. (\ref{3-sigma}) gives the following uncertainty equality:
\begin{equation}
e^{-H_2(\sigma_x)} + e^{-H_2(\sigma_y)} + e^{-H_2(\sigma_z)}
= 1+\mathrm{Tr}[\rho^2]\; ,
\end{equation}
where the monotonic relation $\Delta A^2 = g_2(A)= 2 - 2 e^{-H_2(A)}$ is
employed.

\subsection{Uncertainty relations for spin-$1$ and even higher}

Proposition 1 is generally applicable to arbitrary non-degenerate observables, here we take the spin systems as examples for high dimensional systems. For spin-1 operators $J_a = \vec{J}\cdot \vec{n}_a$ with eigenvalues $\lambda_{1}=1$, $\lambda_{2}=0$, $\lambda_{3}=-1$ (assume $\hbar = 1$) we have
\begin{equation}
\ell_1(J_a)=(J_a^2+J_a)/2\; ,\; \ell_2(J_a) = 1-J_a^2 \; ,\; \ell_3(J_a) = (J_a^2-J_a)/2 \; .
\end{equation}
According to Proposition 1, the covariances of the operators $\ell_2(J_a)$ and $\ell_{3}(J_a)$ can be evaluated and the probability distribution is recovered
\begin{eqnarray}
p_1^2 & = & \frac{\Omega_{12}\Omega_{13}}{\Omega_{23}} = \frac{[V(J_a^2) + \langle J_a^3\rangle - \langle J_a^2\rangle \langle J_a\rangle][V(J_a)-V(J_a^2)]}{4[V(J_a^2) - (\langle J_a^3\rangle - \langle J_a^2\rangle \langle J_a\rangle)]} \; , \\
p_2^2 & = & \frac{\Omega_{12}\Omega_{23}}{\Omega_{13}} =
\frac{V(J_a^2)^2 - (\langle J_a^3\rangle
- \langle J_a^2\rangle \langle J_a\rangle)^2}{V(J_a) -V(J_a^2)} \; , \\
p_3^2 & = & \frac{\Omega_{13}\Omega_{23}}{\Omega_{12}} =
\frac{[V(J_a^2) - (\langle J_a^3\rangle -\langle J_a^2\rangle \langle J_a\rangle)]
[V(J_a) - V(J_a^2)]}{4[V(J_a^2) + \langle J_a^3\rangle -\langle J_a^2\rangle \langle J_a \rangle]} \; .
\end{eqnarray}
The collision entropy now may be expressed as
\begin{eqnarray}
H_2(J_a) & = & -\ln[1-2(p_1p_2+p_1p_3+p_2p_3)] \nonumber \\
& = & -\ln[1-\frac{1}{2}V(J_a)-\frac{3}{2}V(J_a^2)] \; . \label{H2-V-spin1}
\end{eqnarray}
For two operators $J_a$ and $J_b = \vec{J}\cdot \vec{n}_b$, the entropic UR
\begin{equation}
H_2(J_a) + H_2(J_b) \geq c \label{collision-HJ}
\end{equation}
immediately leads to the following variance-based UR
\begin{equation}
[2-V(J_a)-3V(J_a^2)][2-V(J_b)-3V(J_b^2)] \leq 4e^{-c} \; . \label{H2-V-relation}
\end{equation}
Here the lower bound of Eq. (\ref{H2-V-relation}) is optimal if the $c$ in Eq. (\ref{collision-HJ}) is optimal, and the tightness inherits that of Eq. (\ref{collision-HJ}). Pucha{\l}a, {\it et al}. \cite{Majorization-1} had found a simple
bound for Eq. (\ref{collision-HJ}), i.e.
\begin{eqnarray}
c = - \ln[(\frac{1+c_{ab}}{2})^4 +(1-(\frac{1+c_{ab}}{2})^2)^2] \; ,
\label{puchala}
\end{eqnarray}
where $c_{ab}$ is the maximum overlap of eigenbases of operators $J_a$ and
$J_b$. Considering Eq. (\ref{puchala}) for the case of angular momentum operators along the $x$ and $z$ axes, Eq. (\ref{H2-V-relation}) becomes
\begin{equation}
[2-V(J_x)-3V(J_x^2)][2-V(J_z)-3V(J_z^2)]
\leq \frac{25}{8}-\frac{1}{\sqrt{2}} \; . \label{J-example}
\end{equation}
A numerical evaluation of above inequality shows that $V(J_x) + V(J_z)\geq
7/16$, which is consistent with that of Ref. \cite{uncertainty-angular} for spin-1 system. Similar expression as Eq. (\ref{H2-V-spin1}) may also be obtained for spin-$\frac{3}{2}$ system, of which the collision entropy reads
\begin{eqnarray}
H_2(J_a)&  = & - \ln\left\{1-\left[\frac{5}{9}V(J_a^3) + \frac{1}{4}V(J_a^2) + \frac{365}{144}V(J_a) \right.\right. \nonumber \\ & &  \left.\left. -\frac{41}{18}(\langle J_a^4\rangle - \langle J_a\rangle \langle J_a^3\rangle)\right]\right\}\; .
\end{eqnarray}
In principle, there is also no difficult to get similar relations as Eq. (\ref{H2-V-relation}) for even higher spin systems by applying Proposition 1.

To summarize, we have built an one-to-one correspondence between the variance and entropy in qubit system as has been shown in Section 2.2. For high-dimensional systems, different covariance functions are needed to build the probability distributions, see Section 2.3. The measurements of covariance functions involves the measurements of high-order moments of an operator which are compatible with the measurement of its variance (the operator is commuted with its powers). This indicates that the high-order momentum may be a necessity for the further understandings of the entropic and covariance-based uncertainty relations.

\section{Conclusions}

We find in this work an equivalence theorem to unify the
superficially different classes of uncertainty relations, the variance and entropy based ones. For non-degenerate observables, the probability distributions are recovered from the covariance functions of the operators. Among the various applications of this theorem, optimal entropic uncertainty relations containing multiple observables are obtained from the variance based uncertainty relations for qubit system, where when the observables are more than two, the obtained entropic uncertainty relations are in equality form. Explicit functional relations between variance and entropy are constructed for higher spin system. While interest in their own right, these results may also have direct applications in the study of quantum nonlocality, as the uncertainty relations are employed to determine the strength of quantum correlations \cite{correlation-2,correlation-3}. Another important impact of the equivalence theorem is on the structure of the  uncertainty relation in the presence of quantum memory \cite{uncertainty-memory}, which is crucial for the security of quantum key distribution. Finally, since the theorem generally applies to arbitrary dimensional discrete system, it constitutes the basis for further studies of different uncertainty measures and relations.

\section*{Acknowledgments}

\noindent This work was supported in part by Ministry of Science and
Technology of the People's Republic of China (2015CB856703), and by the
National Natural Science Foundation of China (NSFC) under the grants
11175249, 11375200, and 11205239.

\end{document}